\documentstyle[aas2pp4]{article}
\def\PsfigVersion{1.10}
\def\setDriver{\DvipsDriver} % \DvipsDriver or \OzTeXDriver
\ifx\undefined\psfig\else\endinput\fi
\let\LaTeXAtSign=\@
\let\@=\relax
\edef\psfigRestoreAt{\catcode`\@=\number\catcode`@\relax}
\catcode`\@=11\relax
\newwrite\@unused
\def\ps@typeout#1{{\let\protect\string\immediate\write\@unused{#1}}}

\def\DvipsDriver{
	\ps@typeout{psfig/tex \PsfigVersion -dvips}
\def\PsfigSpecials{\DvipsSpecials} 	\def\ps@dir{/}
\def\ps@predir{} }
\def\OzTeXDriver{
	\ps@typeout{psfig/tex \PsfigVersion -oztex}
	\def\PsfigSpecials{\OzTeXSpecials}
	\def\ps@dir{:}
	\def\ps@predir{:}
	\catcode`\^^J=5
}

\def\figurepath{./:}

\def\DoPaths#1{\expandafter\EachPath#1\stoplist}
\def\leer{}
\def\EachPath#1:#2\stoplist{% #1 part of the list (delimiter :)
  \ExistsFile{#1}{\SearchedFile}
  \ifx#2\leer
  \else
    \expandafter\EachPath#2\stoplist
  \fi}
\def\ps@dir{/}
\def\ExistsFile#1#2{%
   \openin1=\ps@predir#1\ps@dir#2
   \ifeof1
       \closein1
   \else
       \closein1
        \ifx\ps@founddir\leer
           \edef\ps@founddir{#1}
        \fi
   \fi}
\def\get@dir#1{%
  \def\ps@founddir{}
  \def\SearchedFile{#1}
  \DoPaths\figurepath
}
\def\@nnil{\@nil}
\def\@empty{}
\def\@psdonoop#1\@@#2#3{}
\def\@psdo#1:=#2\do#3{\edef\@psdotmp{#2}\ifx\@psdotmp\@empty \else
    \expandafter\@psdoloop#2,\@nil,\@nil\@@#1{#3}\fi}
\def\@psdoloop#1,#2,#3\@@#4#5{\def#4{#1}\ifx #4\@nnil \else
       #5\def#4{#2}\ifx #4\@nnil \else#5\@ipsdoloop #3\@@#4{#5}\fi\fi}
\def\@ipsdoloop#1,#2\@@#3#4{\def#3{#1}\ifx #3\@nnil 
       \let\@nextwhile=\@psdonoop \else
      #4\relax\let\@nextwhile=\@ipsdoloop\fi\@nextwhile#2\@@#3{#4}}
\def\@tpsdo#1:=#2\do#3{\xdef\@psdotmp{#2}\ifx\@psdotmp\@empty \else
    \@tpsdoloop#2\@nil\@nil\@@#1{#3}\fi}
\def\@tpsdoloop#1#2\@@#3#4{\def#3{#1}\ifx #3\@nnil 
       \let\@nextwhile=\@psdonoop \else
      #4\relax\let\@nextwhile=\@tpsdoloop\fi\@nextwhile#2\@@#3{#4}}
\ifx\undefined\fbox
\newdimen\fboxrule
\newdimen\fboxsep
\newdimen\ps@tempdima
\newbox\ps@tempboxa
\long\def\fbox#1{\leavevmode\setbox\ps@tempboxa\hbox{#1}\ps@tempdima\fboxrule
    \advance\ps@tempdima \fboxsep \advance\ps@tempdima \dp\ps@tempboxa
   \hbox{\lower \ps@tempdima\hbox
  {\vbox{\hrule height \fboxrule
          \hbox{\vrule width \fboxrule \hskip\fboxsep
          \vbox{\vskip\fboxsep \box\ps@tempboxa\vskip\fboxsep}\hskip 
                 \fboxsep\vrule width \fboxrule}
                 \hrule height \fboxrule}}}}
\fi
\newread\ps@stream
\newif\ifnot@eof       % continue looking for the bounding box?
\newif\if@noisy        % report what you're making?
\newif\if@atend        % %%BoundingBox: has (at end) specification
\newif\if@psfile       % does this look like a PostScript file?
{\catcode`\%=12\global\gdef\epsf@start{%!}}
\def\epsf@PS{PS}
\def\epsf@getbb#1{%
\openin\ps@stream=\ps@predir#1
\ifeof\ps@stream\ps@typeout{Error, File #1 not found}\else
   {\not@eoftrue \chardef\other=12
    \def\do##1{\catcode`##1=\other}\dospecials \catcode`\ =10
    \loop
       \if@psfile
	  \read\ps@stream to \epsf@fileline
       \else{
	  \obeyspaces
          \read\ps@stream to \epsf@tmp\global\let\epsf@fileline\epsf@tmp}
       \fi
       \ifeof\ps@stream\not@eoffalse\else
       \if@psfile\else
       \expandafter\epsf@test\epsf@fileline:. \\%
       \fi
          \expandafter\epsf@aux\epsf@fileline:. \\%
       \fi
   \ifnot@eof\repeat
   }\closein\ps@stream\fi}%
\long\def\epsf@test#1#2#3:#4\\{\def\epsf@testit{#1#2}
			\ifx\epsf@testit\epsf@start\else
\ps@typeout{Warning! File does not start with `\epsf@start'.  It may not be a PostScript file.}
			\fi
			\@psfiletrue} % don't test after 1st line
{\catcode`\%=12\global\let\epsf@percent=%\global\def\epsf@bblit{%BoundingBox}}
\long\def\epsf@aux#1#2:#3\\{\ifx#1\epsf@percent
   \def\epsf@testit{#2}\ifx\epsf@testit\epsf@bblit
	\@atendfalse
        \epsf@atend #3 . \\%
	\if@atend	
	   \if@verbose{
		\ps@typeout{psfig: found `(atend)'; continuing search}
	   }\fi
        \else
        \epsf@grab #3 . . . \\%
        \not@eoffalse
        \global\no@bbfalse
        \fi
   \fi\fi}%
\def\epsf@grab #1 #2 #3 #4 #5\\{%
   \global\def\epsf@llx{#1}\ifx\epsf@llx\empty
      \epsf@grab #2 #3 #4 #5 .\\\else
   \global\def\epsf@lly{#2}%
   \global\def\epsf@urx{#3}\global\def\epsf@ury{#4}\fi}%
\def\epsf@atendlit{(atend)} 
\def\epsf@atend #1 #2 #3\\{%
   \def\epsf@tmp{#1}\ifx\epsf@tmp\empty
      \epsf@atend #2 #3 .\\\else
   \ifx\epsf@tmp\epsf@atendlit\@atendtrue\fi\fi}

\chardef\psletter = 11 % won't conflict with \begin{letter} now...
\chardef\other = 12

\newif \ifdebug %%% turn me on to see TeX hard at work ...
\newif\ifc@mpute %%% don't need to compute some values
\c@mputetrue % but assume that we do

\let\then = \relax
\def\r@dian{pt }
\let\r@dians = \r@dian
\let\dimensionless@nit = \r@dian
\let\dimensionless@nits = \dimensionless@nit
\def\internal@nit{sp }
\let\internal@nits = \internal@nit
\newif\ifstillc@nverging
\def \Mess@ge #1{\ifdebug \then \message {#1} \fi}

{ %%% Things that need abnormal catcodes %%%
	\catcode `\@ = \psletter
	\gdef \nodimen {\expandafter \n@dimen \the \dimen}
	\gdef \term #1 #2 #3%
	       {\edef \t@ {\the #1}%%% freeze parameter 1 (count, by value)
		\edef \t@@ {\expandafter \n@dimen \the #2\r@dian}%
		\t@rm {\t@} {\t@@} {#3}%
	       }
	\gdef \t@rm #1 #2 #3%
	       {{%
		\count 0 = 0
		\dimen 0 = 1 \dimensionless@nit
		\dimen 2 = #2\relax
		\Mess@ge {Calculating term #1 of \nodimen 2}%
		\loop
		\ifnum	\count 0 < #1
		\then	\advance \count 0 by 1
			\Mess@ge {Iteration \the \count 0 \space}%
			\Multiply \dimen 0 by {\dimen 2}%
			\Mess@ge {After multiplication, term = \nodimen 0}%
			\Divide \dimen 0 by {\count 0}%
			\Mess@ge {After division, term = \nodimen 0}%
		\repeat
		\Mess@ge {Final value for term #1 of 
				\nodimen 2 \space is \nodimen 0}%
		\xdef \Term {#3 = \nodimen 0 \r@dians}%
		\aftergroup \Term
	       }}
	\catcode `\p = \other
	\catcode `\t = \other
	\gdef \n@dimen #1pt{#1} %%% throw away the ``pt''
}

\def \Divide #1by #2{\divide #1 by #2} %%% just a synonym

\def \Multiply #1by #2%%% allows division of a dimen by a dimen
       {{%%% should really freeze parameter 2 (dimen, passed by value)
	\count 0 = #1\relax
	\count 2 = #2\relax
	\count 4 = 65536
	\Mess@ge {Before scaling, count 0 = \the \count 0 \space and
			count 2 = \the \count 2}%
	\ifnum	\count 0 > 32767 %%% do our best to avoid overflow
	\then	\divide \count 0 by 4
		\divide \count 4 by 4
	\else	\ifnum	\count 0 < -32767
		\then	\divide \count 0 by 4
			\divide \count 4 by 4
		\else
		\fi
	\fi
	\ifnum	\count 2 > 32767 %%% while retaining reasonable accuracy
	\then	\divide \count 2 by 4
		\divide \count 4 by 4
	\else	\ifnum	\count 2 < -32767
		\then	\divide \count 2 by 4
			\divide \count 4 by 4
		\else
		\fi
	\fi
	\multiply \count 0 by \count 2
	\divide \count 0 by \count 4
	\xdef \product {#1 = \the \count 0 \internal@nits}%
	\aftergroup \product
       }}

\def\r@duce{\ifdim\dimen0 > 90\r@dian \then   % sin(x+90) = sin(180-x)
		\multiply\dimen0 by -1
		\advance\dimen0 by 180\r@dian
		\r@duce
	    \else \ifdim\dimen0 < -90\r@dian \then  % sin(-x) = sin(360+x)
		\advance\dimen0 by 360\r@dian
		\r@duce
		\fi
	    \fi}

\def\Sine#1%
       {{%
	\dimen 0 = #1 \r@dian
	\r@duce
	\ifdim\dimen0 = -90\r@dian \then
	   \dimen4 = -1\r@dian
	   \c@mputefalse
	\fi
	\ifdim\dimen0 = 90\r@dian \then
	   \dimen4 = 1\r@dian
	   \c@mputefalse
	\fi
	\ifdim\dimen0 = 0\r@dian \then
	   \dimen4 = 0\r@dian
	   \c@mputefalse
	\fi
	\ifc@mpute \then
		\divide\dimen0 by 180
		\dimen0=3.141592654\dimen0
		\dimen 2 = 3.1415926535897963\r@dian %%% a well-known constant
		\divide\dimen 2 by 2 %%% we only deal with -pi/2 : pi/2
		\Mess@ge {Sin: calculating Sin of \nodimen 0}%
		\count 0 = 1 %%% see power-series expansion for sine
		\dimen 2 = 1 \r@dian %%% ditto
		\dimen 4 = 0 \r@dian %%% ditto
		\loop
			\ifnum	\dimen 2 = 0 %%% then we've done
			\then	\stillc@nvergingfalse 
			\else	\stillc@nvergingtrue
			\fi
			\ifstillc@nverging %%% then calculate next term
			\then	\term {\count 0} {\dimen 0} {\dimen 2}%
				\advance \count 0 by 2
				\count 2 = \count 0
				\divide \count 2 by 2
				\ifodd	\count 2 %%% signs alternate
				\then	\advance \dimen 4 by \dimen 2
				\else	\advance \dimen 4 by -\dimen 2
				\fi
		\repeat
	\fi		
			\xdef \sine {\nodimen 4}%
       }}

\def\Cosine#1{\ifx\sine\UnDefined\edef\Savesine{\relax}\else
		             \edef\Savesine{\sine}\fi
	{\dimen0=#1\r@dian\advance\dimen0 by 90\r@dian
	 \Sine{\nodimen 0}
	 \xdef\cosine{\sine}
	 \xdef\sine{\Savesine}}}	      
\def\psdraft{
	\def\@psdraft{0}
}
\def\psfull{
	\def\@psdraft{100}
}

\psfull

\newif\if@scalefirst
\def\psscalefirst{\@scalefirsttrue}
\def\psrotatefirst{\@scalefirstfalse}
\psrotatefirst

\newif\if@draftbox
\def\psnodraftbox{
	\@draftboxfalse
}
\def\psdraftbox{
	\@draftboxtrue
}
\@draftboxtrue

\newif\if@prologfile
\newif\if@postlogfile
\def\pssilent{
	\@noisyfalse
}
\def\psnoisy{
	\@noisytrue
}
\psnoisy
\newif\if@bbllx
\newif\if@bblly
\newif\if@bburx
\newif\if@bbury
\newif\if@height
\newif\if@width
\newif\if@rheight
\newif\if@rwidth
\newif\if@angle
\newif\if@clip
\newif\if@verbose
\def\@p@@sclip#1{\@cliptrue}
\newif\if@decmpr
\def\@p@@sfigure#1{\def\@p@sfile{null}\def\@p@sbbfile{null}\@decmprfalse
   \openin1=\ps@predir#1
   \ifeof1
	\closein1
	\get@dir{#1}
	\ifx\ps@founddir\leer
		\openin1=\ps@predir#1.bb
		\ifeof1
			\closein1
			\get@dir{#1.bb}
			\ifx\ps@founddir\leer
				\ps@typeout{Can't find #1 in \figurepath}
			\else
				\@decmprtrue
				\def\@p@sfile{\ps@founddir\ps@dir#1}
				\def\@p@sbbfile{\ps@founddir\ps@dir#1.bb}
			\fi
		\else
			\closein1
			\@decmprtrue
			\def\@p@sfile{#1}
			\def\@p@sbbfile{#1.bb}
		\fi
	\else
		\def\@p@sfile{\ps@founddir\ps@dir#1}
		\def\@p@sbbfile{\ps@founddir\ps@dir#1}
	\fi
   \else
	\closein1
	\def\@p@sfile{#1}
	\def\@p@sbbfile{#1}
   \fi
}
\def\@p@@sfile#1{\@p@@sfigure{#1}}
\def\@p@@sbbllx#1{
		\@bbllxtrue
		\dimen100=#1
		\edef\@p@sbbllx{\number\dimen100}
}
\def\@p@@sbblly#1{
		\@bbllytrue
		\dimen100=#1
		\edef\@p@sbblly{\number\dimen100}
}
\def\@p@@sbburx#1{
		\@bburxtrue
		\dimen100=#1
		\edef\@p@sbburx{\number\dimen100}
}
\def\@p@@sbbury#1{
		\@bburytrue
		\dimen100=#1
		\edef\@p@sbbury{\number\dimen100}
}
\def\@p@@sheight#1{
		\@heighttrue
		\dimen100=#1
   		\edef\@p@sheight{\number\dimen100}
}
\def\@p@@swidth#1{
		\@widthtrue
		\dimen100=#1
		\edef\@p@swidth{\number\dimen100}
}
\def\@p@@srheight#1{
		\@rheighttrue
		\dimen100=#1
		\edef\@p@srheight{\number\dimen100}
}
\def\@p@@srwidth#1{
		\@rwidthtrue
		\dimen100=#1
		\edef\@p@srwidth{\number\dimen100}
}
\def\@p@@sangle#1{
		\@angletrue
		\edef\@p@sangle{#1} %\number\dimen100}
}
\def\@p@@ssilent#1{ 
		\@verbosefalse
}
\def\@p@@sprolog#1{\@prologfiletrue\def\@prologfileval{#1}}
\def\@p@@spostlog#1{\@postlogfiletrue\def\@postlogfileval{#1}}
\def\@cs@name#1{\csname #1\endcsname}
\def\@setparms#1=#2,{\@cs@name{@p@@s#1}{#2}}
\def\ps@init@parms{
		\@bbllxfalse \@bbllyfalse
		\@bburxfalse \@bburyfalse
		\@heightfalse \@widthfalse
		\@rheightfalse \@rwidthfalse
		\def\@p@sbbllx{}\def\@p@sbblly{}
		\def\@p@sbburx{}\def\@p@sbbury{}
		\def\@p@sheight{}\def\@p@swidth{}
		\def\@p@srheight{}\def\@p@srwidth{}
		\def\@p@sangle{0}
		\def\@p@sfile{} \def\@p@sbbfile{}
		\def\@p@scost{10}
		\def\@sc{}
		\@prologfilefalse
		\@postlogfilefalse
		\@clipfalse
		\if@noisy
			\@verbosetrue
		\else
			\@verbosefalse
		\fi
}
\def\parse@ps@parms#1{
	 	\@psdo\@psfiga:=#1\do
		   {\expandafter\@setparms\@psfiga,}}
\newif\ifno@bb
\def\bb@missing{
	\if@verbose{
		\ps@typeout{psfig: searching \@p@sbbfile \space  for bounding box}
	}\fi
	\no@bbtrue
	\epsf@getbb{\@p@sbbfile}
        \ifno@bb \else \bb@cull\epsf@llx\epsf@lly\epsf@urx\epsf@ury\fi
}	
\def\bb@cull#1#2#3#4{
	\dimen100=#1 bp\edef\@p@sbbllx{\number\dimen100}
	\dimen100=#2 bp\edef\@p@sbblly{\number\dimen100}
	\dimen100=#3 bp\edef\@p@sbburx{\number\dimen100}
	\dimen100=#4 bp\edef\@p@sbbury{\number\dimen100}
	\no@bbfalse
}
\newdimen\p@intvaluex
\newdimen\p@intvaluey
\def\rotate@#1#2{{\dimen0=#1 sp\dimen1=#2 sp
		  \global\p@intvaluex=\cosine\dimen0
		  \dimen3=\sine\dimen1
		  \global\advance\p@intvaluex by -\dimen3
		  \global\p@intvaluey=\sine\dimen0
		  \dimen3=\cosine\dimen1
		  \global\advance\p@intvaluey by \dimen3
		  }}
\def\compute@bb{
		\no@bbfalse
		\if@bbllx \else \no@bbtrue \fi
		\if@bblly \else \no@bbtrue \fi
		\if@bburx \else \no@bbtrue \fi
		\if@bbury \else \no@bbtrue \fi
		\ifno@bb \bb@missing \fi
		\ifno@bb \ps@typeout{FATAL ERROR: no bb supplied or found}
			\no-bb-error
		\fi
		\count203=\@p@sbburx
		\count204=\@p@sbbury
		\advance\count203 by -\@p@sbbllx
		\advance\count204 by -\@p@sbblly
		\edef\ps@bbw{\number\count203}
		\edef\ps@bbh{\number\count204}
		\if@angle 
			\Sine{\@p@sangle}\Cosine{\@p@sangle}
	        	{\dimen100=\maxdimen\xdef\r@p@sbbllx{\number\dimen100}
					    \xdef\r@p@sbblly{\number\dimen100}
			                    \xdef\r@p@sbburx{-\number\dimen100}
					    \xdef\r@p@sbbury{-\number\dimen100}}
                        \def\minmaxtest{
			   \ifnum\number\p@intvaluex<\r@p@sbbllx
			      \xdef\r@p@sbbllx{\number\p@intvaluex}\fi
			   \ifnum\number\p@intvaluex>\r@p@sbburx
			      \xdef\r@p@sbburx{\number\p@intvaluex}\fi
			   \ifnum\number\p@intvaluey<\r@p@sbblly
			      \xdef\r@p@sbblly{\number\p@intvaluey}\fi
			   \ifnum\number\p@intvaluey>\r@p@sbbury
			      \xdef\r@p@sbbury{\number\p@intvaluey}\fi
			   }
			\rotate@{\@p@sbbllx}{\@p@sbblly}
			\minmaxtest
			\rotate@{\@p@sbbllx}{\@p@sbbury}
			\minmaxtest
			\rotate@{\@p@sbburx}{\@p@sbblly}
			\minmaxtest
			\rotate@{\@p@sbburx}{\@p@sbbury}
			\minmaxtest
			\edef\@p@sbbllx{\r@p@sbbllx}\edef\@p@sbblly{\r@p@sbblly}
			\edef\@p@sbburx{\r@p@sbburx}\edef\@p@sbbury{\r@p@sbbury}
		\fi
		\count203=\@p@sbburx
		\count204=\@p@sbbury
		\advance\count203 by -\@p@sbbllx
		\advance\count204 by -\@p@sbblly
		\edef\@bbw{\number\count203}
		\edef\@bbh{\number\count204}
}
\def\in@hundreds#1#2#3{\count240=#2 \count241=#3
		     \count100=\count240	% 100 is first digit #2/#3
		     \divide\count100 by \count241
		     \count101=\count100
		     \multiply\count101 by \count241
		     \advance\count240 by -\count101
		     \multiply\count240 by 10
		     \count101=\count240	%101 is second digit of #2/#3
		     \divide\count101 by \count241
		     \count102=\count101
		     \multiply\count102 by \count241
		     \advance\count240 by -\count102
		     \multiply\count240 by 10
		     \count102=\count240	% 102 is the third digit
		     \divide\count102 by \count241
		     \count200=#1\count205=0
		     \count201=\count200
			\multiply\count201 by \count100
		 	\advance\count205 by \count201
		     \count201=\count200
			\divide\count201 by 10
			\multiply\count201 by \count101
			\advance\count205 by \count201
		     \count201=\count200
			\divide\count201 by 100
			\multiply\count201 by \count102
			\advance\count205 by \count201
		     \edef\@result{\number\count205}
}
\def\compute@wfromh{
		\in@hundreds{\@p@sheight}{\@bbw}{\@bbh}
		\edef\@p@swidth{\@result}
}
\def\compute@hfromw{
	        \in@hundreds{\@p@swidth}{\@bbh}{\@bbw}
		\edef\@p@sheight{\@result}
}
\def\compute@handw{
		\if@height 
			\if@width
			\else
				\compute@wfromh
			\fi
		\else 
			\if@width
				\compute@hfromw
			\else
				\edef\@p@sheight{\@bbh}
				\edef\@p@swidth{\@bbw}
			\fi
		\fi
}
\def\compute@resv{
		\if@rheight \else \edef\@p@srheight{\@p@sheight} \fi
		\if@rwidth \else \edef\@p@srwidth{\@p@swidth} \fi
}
\def\compute@sizes{
	\compute@bb
	\if@scalefirst\if@angle
	\if@width
	   \in@hundreds{\@p@swidth}{\@bbw}{\ps@bbw}
	   \edef\@p@swidth{\@result}
	\fi
	\if@height
	   \in@hundreds{\@p@sheight}{\@bbh}{\ps@bbh}
	   \edef\@p@sheight{\@result}
	\fi
	\fi\fi
	\compute@handw
	\compute@resv}
\def\OzTeXSpecials{
	\special{empty.ps /@isp {true} def}
	\special{empty.ps \@p@swidth \space \@p@sheight \space
			\@p@sbbllx \space \@p@sbblly \space
			\@p@sbburx \space \@p@sbbury \space
			startTexFig \space }
	\if@clip{
		\if@verbose{
			\ps@typeout{(clip)}
		}\fi
		\special{empty.ps doclip \space }
	}\fi
	\if@angle{
		\if@verbose{
			\ps@typeout{(rotate)}
		}\fi
		\special {empty.ps \@p@sangle \space rotate \space} 
	}\fi
	\if@prologfile
	    \special{\@prologfileval \space } \fi
	\if@decmpr{
		\if@verbose{
			\ps@typeout{psfig: Compression not available
			in OzTeX version \space }
		}\fi
	}\else{
		\if@verbose{
			\ps@typeout{psfig: including \@p@sfile \space }
		}\fi
		\special{epsf=\ps@predir\@p@sfile \space }
	}\fi
	\if@postlogfile
	    \special{\@postlogfileval \space } \fi
	\special{empty.ps /@isp {false} def}
}
\def\DvipsSpecials{
	\special{ps::[begin] 	\@p@swidth \space \@p@sheight \space
			\@p@sbbllx \space \@p@sbblly \space
			\@p@sbburx \space \@p@sbbury \space
			startTexFig \space }
	\if@clip{
		\if@verbose{
			\ps@typeout{(clip)}
		}\fi
		\special{ps:: doclip \space }
	}\fi
	\if@angle
		\if@verbose{
			\ps@typeout{(clip)}
		}\fi
		\special {ps:: \@p@sangle \space rotate \space} 
	\fi
	\if@prologfile
	    \special{ps: plotfile \@prologfileval \space } \fi
	\if@decmpr{
		\if@verbose{
			\ps@typeout{psfig: including \@p@sfile.Z \space }
		}\fi
		\special{ps: plotfile "`zcat \@p@sfile.Z" \space }
	}\else{
		\if@verbose{
			\ps@typeout{psfig: including \@p@sfile \space }
		}\fi
		\special{ps: plotfile \@p@sfile \space }
	}\fi
	\if@postlogfile
	    \special{ps: plotfile \@postlogfileval \space } \fi
	\special{ps::[end] endTexFig \space }
}
\def\psfig#1{\vbox {
	\ps@init@parms
	\parse@ps@parms{#1}
	\compute@sizes
	\ifnum\@p@scost<\@psdraft{
		\PsfigSpecials 
		\vbox to \@p@srheight sp{
			\hbox to \@p@srwidth sp{
				\hss
			}
		\vss
		}
	}\else{
		\if@draftbox{		
			\hbox{\fbox{\vbox to \@p@srheight sp{
			\vss
			\hbox to \@p@srwidth sp{ \hss 
			 \hss }
			\vss
			}}}
		}\else{
			\vbox to \@p@srheight sp{
			\vss
			\hbox to \@p@srwidth sp{\hss}
			\vss
			}
		}\fi

	}\fi
}}
\psfigRestoreAt
\setDriver
\let\@=\LaTeXAtSign

\def\Ha{H$\alpha$\,}
\def\Hb{H$\beta$\,}
\def\Msun{M_\odot}
\def\kms{$\rm{km\,s^{-1}}$}
\def\arcsec{$^{\prime\prime}$}
\def\ergs{erg\,s$^{-1}$}
\def\Po{$P_{\rm o}$\ }
\def\degree{$^{\rm o}$}
\baselineskip=12pt plus 2pt 
\begin{document}

\title
{
Polarimetry of the Type Ia Supernova SN 1996X.
}

\author{Lifan Wang\footnote{Also Beijing Astronomical Observatory, 
Beijing 100080, P. R. China}$^,$\footnote{email: lifan@astro.as.utexas.edu}, 
J. Craig Wheeler\footnote{email: 
wheel@astro.as.utexas.edu},
Peter H\"oflich\footnote{email: pah@alla.as.utexas.edu}
\footnote{Also: Institute for Theoretical Physics, University of Basel,
Klingelbergstr. 82, CH-4056 Basel, Switzerland}}

\affil{Department of Astronomy and McDonald Observatory\\
          The University of Texas at Austin\\
          Austin,~TX~78712}

\begin{abstract}
We present broad-band and spectropolarimetry of the Type Ia SN 1996X
obtained on April 14, 1996 (UT), and broad-band polarimetry of SN 1996X on 
May 22, 1996 (UT), when the supernova was about a week before 
and 4 weeks after optical maximum, respectively. The Stokes parameters 
derived from the broad-band polarimetry are consistent with zero 
polarization. The spectropolarimetry, however, shows broad spectral 
features which are due intrinsically to an asymmetric supernova 
atmosphere. The spectral features in the flux spectrum and the 
polarization spectrum show correlations in the wavelength range 
from 4900 \AA\ up to 5500 \AA. The degree of this intrinsic component is 
low ($\sim 0.3\%$). Theoretical polarization spectra have been calculated. 
It is shown that the polarization spectra are governed by line blending.
Consequently, for similar geometrical distortions, the residual polarization 
is smaller by about a factor of 2 to 3 compared to the less blended Type II 
atmosphere, making it intrinsically harder to detect asphericities in SNIa.
Comparison with theoretical model polarization spectra shows a resemblance to 
the observations. Taken literally, this implies an asphericity
of $\approx $ 11 \% in the chemical distribution in the region of partial
burning. This may not imperil the use of Type Ia supernovae 
as standard candles for distance determination, but nontheless poses a 
source of uncertainty. SN 1996X is the first Type Ia supernova for 
which spectropolarimetry revealed a polarized component intrinsic to
the supernova and the first Type Ia with spectropolarimetry well prior 
to optical maximum.  
 
\end{abstract}

\keywords{stars: individual (SN 1996X) -- stars: supernovae -- stars: 
polarimetry} 

\section{Introduction}

Polarimetry is a powerful way of probing asymmetries in supernova explosions. 
Because polarization produced within the supernova atmosphere may have
spectral features that are distinguishable from the featureless interstellar
polarization (ISP), which is caused by dust within both the Galaxy and the 
host galaxy, spectropolarimetry is usually a more powerful tool than 
broad-band polarimetry in decomposing the intrinsic polarization of a distant 
supernova from ISP.

SN 1987A and SN 1993J are the only two supernovae for which spectropolarimetry 
has been published and intrinsic polarizations are positively detected. 
SN 1987A is so far the best observed supernova for which linear polarimetry
was obtained from several days to about 260 days after explosion 
(Mendez et al. 1988; Cropper et al. 1988; Jeffery 1991a). The degree of 
polarization evolved with time, indicating that the cause of the polarization 
is related intrinsically to SN 1987A. The data were analyzed (H\"oflich 1987; 
Jeffery 1991b) in terms of the photospheric scattering model 
(Brown \& McLean 1977; Shapiro \& Sutherland 1982). Recently, 
Wang \& Wheeler (1996a) provided a different view in which time delayed 
scattering by a hypothesized circumstellar dust clump successfully reproduced 
both the broad-band and spectropolarimetry of the supernova and the early 
infrared light curve of SN 1987A (Bouchet et al. 1989). Linear polarization 
indicated by polarization changes across spectral features was also detected 
in SN 1993J (Trammell, Hines, \& Wheeler 1993). Broad band polarimetry shows 
also variable polarization before and after the second optical maximum of 
SN 1993J (Doroshenko, Efimov, \& Shakhovskoi 1995).

Only two Type Ia supernovae -- SN 1983G (McCall et al. 1984) and SN 1992A 
(Spyromilio \& Bailey 1992) have been previously observed by spectropolarimetry.
The noise levels of the SN 1983G and SN 1992A data are 0.5\% and 0.3\%, 
respectively. These data indicate SN 1983G and SN 1992A are not polarized at 
a level higher than 0.5\%.

As part of our program of systematic supernova polarimetry, we have observed
5 supernovae with broad-band polarimetry and compiled a catalog of all the
supernovae with polarimetry reported in the literature (Wang et al. 1996). 
This sample shows that all the Type II supernovae with sufficient data are 
intrinsically polarized while no intrinsic polarization can be established 
for any of the Type Ia supernovae.

Supernovae, of both Type Ia and II, are now being widely used as standard
candles for distance determinations. As a self-consistency test,
it is imperative to set some observational constraints on the degree of 
the polarization. Furthermore, normal spectroscopy provides only part of 
the information carried by the supernova light. Spectropolarimetry can 
further constrain various models for spectral line formation and radiative 
transfer through supernova atmospheres, as shown in recent examples by
H\"oflich (1995a), H\"oflich et al. (1996), and H\"oflich, Wheeler, \& Wang 
(1997). Furthermore, it may be the only way to detect chemical blobs in the
ejecta which usually produce little observable effects on the flux spectra
of a supernova (H\"oflich, Wheeler, \& Wang 1997).

\section{The Observations}

SN 1996X in NGC 5061 was discovered independently by several observers 
(Garradd 1996) on April 12.5 UT at a V magnitude about 13.5. Spectroscopy 
on April 14 UT showed SN 1996X to be a Type Ia supernova prior to optical 
maximum (Suntzeff 1996; Wang and Wheeler 1996b; Benetti and Patat 1996). 
Our spectropolarimetry of SN 1996X was obtained on April 14.3 using the 
Imaging Grism Polarimeter (IGP) mounted at the Cassegrain focus of the 
2.1 meter telescope of the McDonald Observatory. The IGP is a 
simple, high efficiency, dual beam polarimeter which can be easily switched 
between imaging and spectropolarimetry mode. A rotatable half waveplate was 
used as the polarization analyzer. A description of the instrument can be 
found in Hill \& Trammell (1996) and Trammell (1994). A TK4 ($1024\times1024$) 
CCD was used as the detector. A broad filter with pass band from 4400 to 7500 
was used for the broad-band polarimetry. Such a filter effectively filters 
out photons with wavelengths beyond the effective range of the waveplate, but 
allows a large number of photons to be transmitted for polarimetry. The slit 
width used for the spectropolarimetry was 2\arcsec.1 which gave a spectral 
resolution of about $14$\AA. Wavelength calibration was obtained by taking 
exposures of an argon lamp. 

\begin{figure}
\psfig{figure=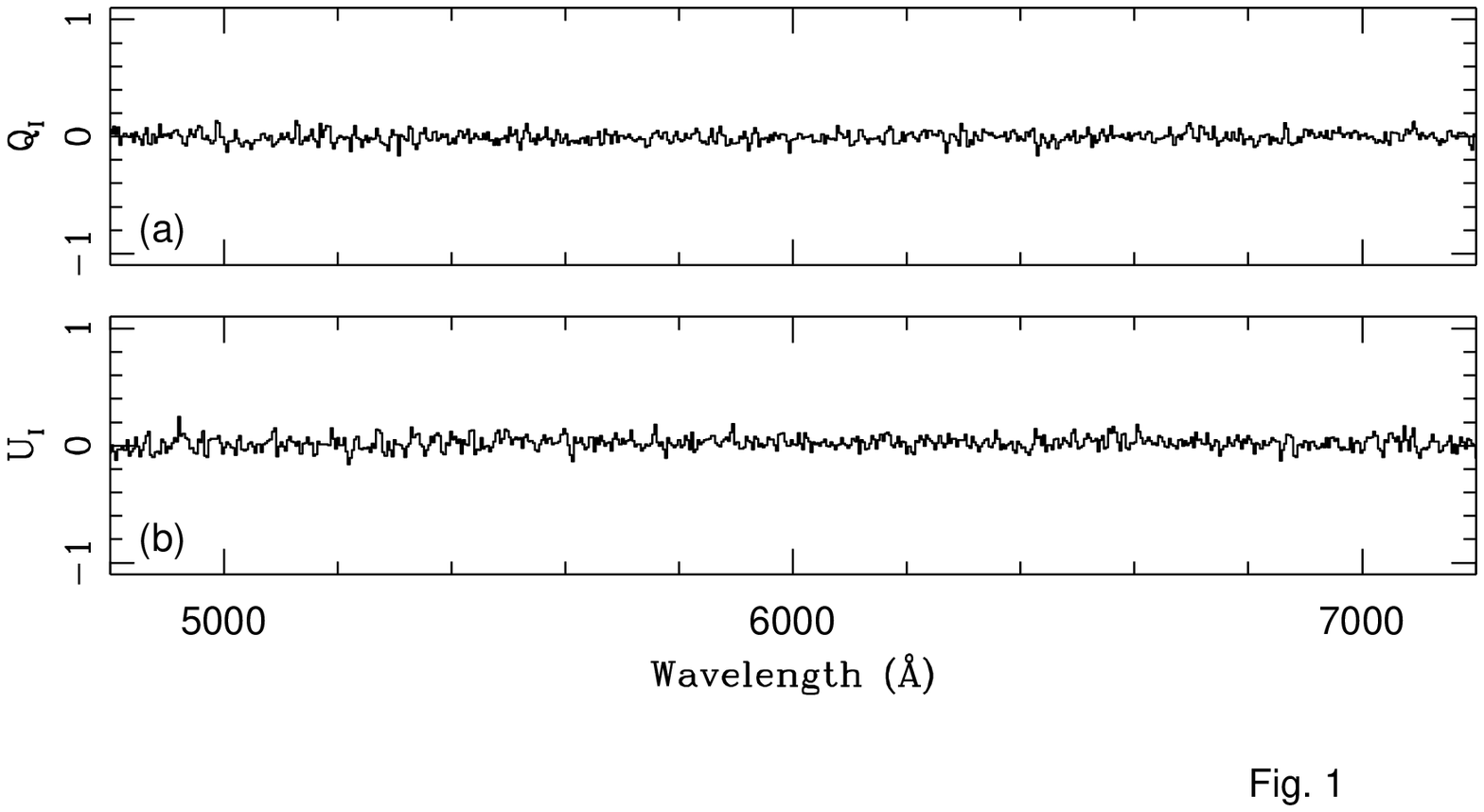,width=7.8cm,clip=,angle=0}
 \caption{ Instrumental polarization from observations of unpolarized 
             standard stars. }
\psfig{figure=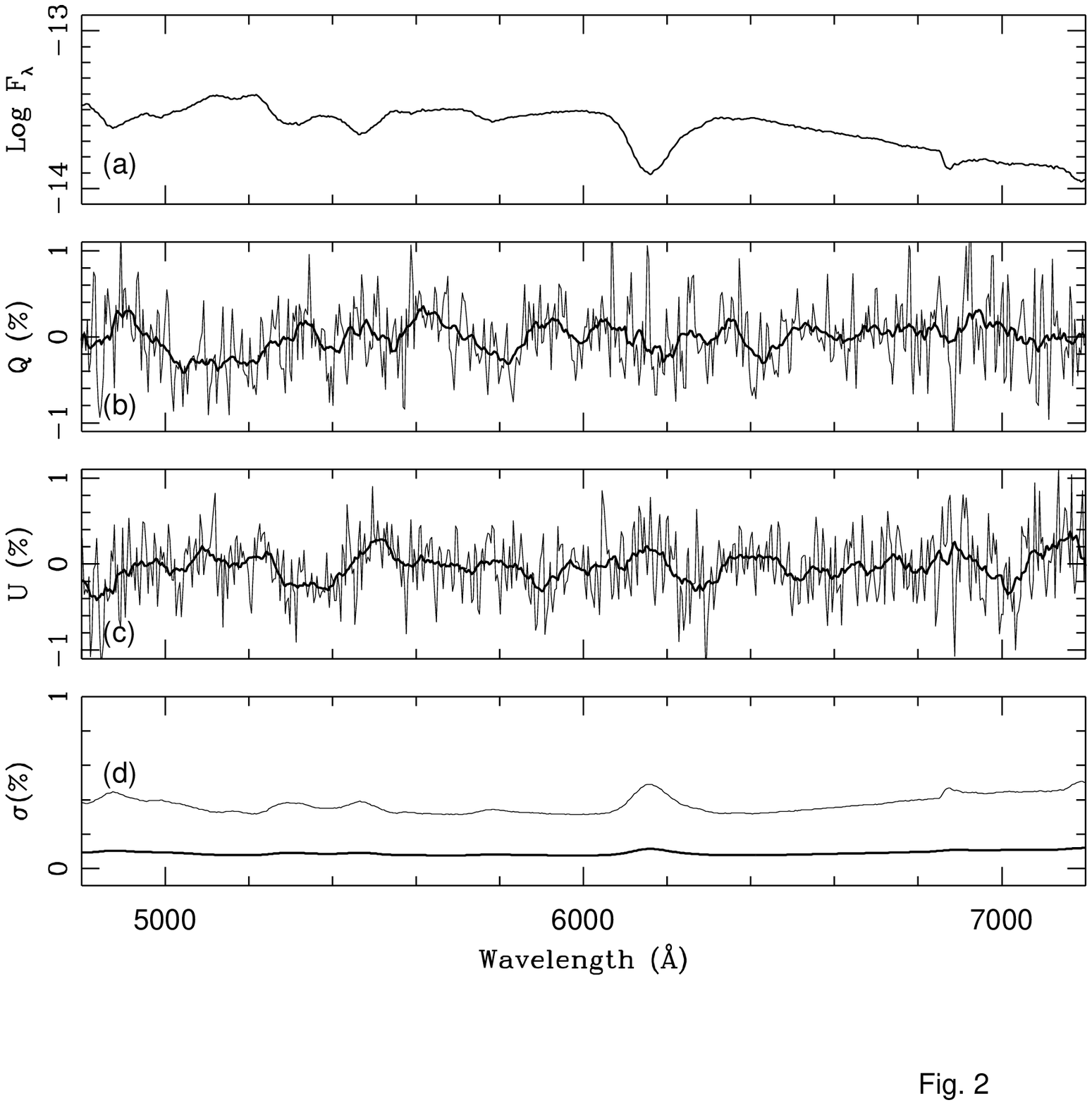,width=7.8cm,angle=0}
 \caption{
 (a) The flux spectra of SN 1996X on April 14.3, 1996 UT in 
             arbitrary units; (b) the Stokes parameter $Q$; (c) the Stokes 
             parameter
             $U$; (d) the noise level of the Stokes parameters. In (b) - (c), 
             the thin lines are the observed data at the original sampling
             step of 3.8\AA/pixel, the thick lines are the same data 
             resampled with a sample window width of 64.8\AA (17 pixels).   
}
\psfig{figure=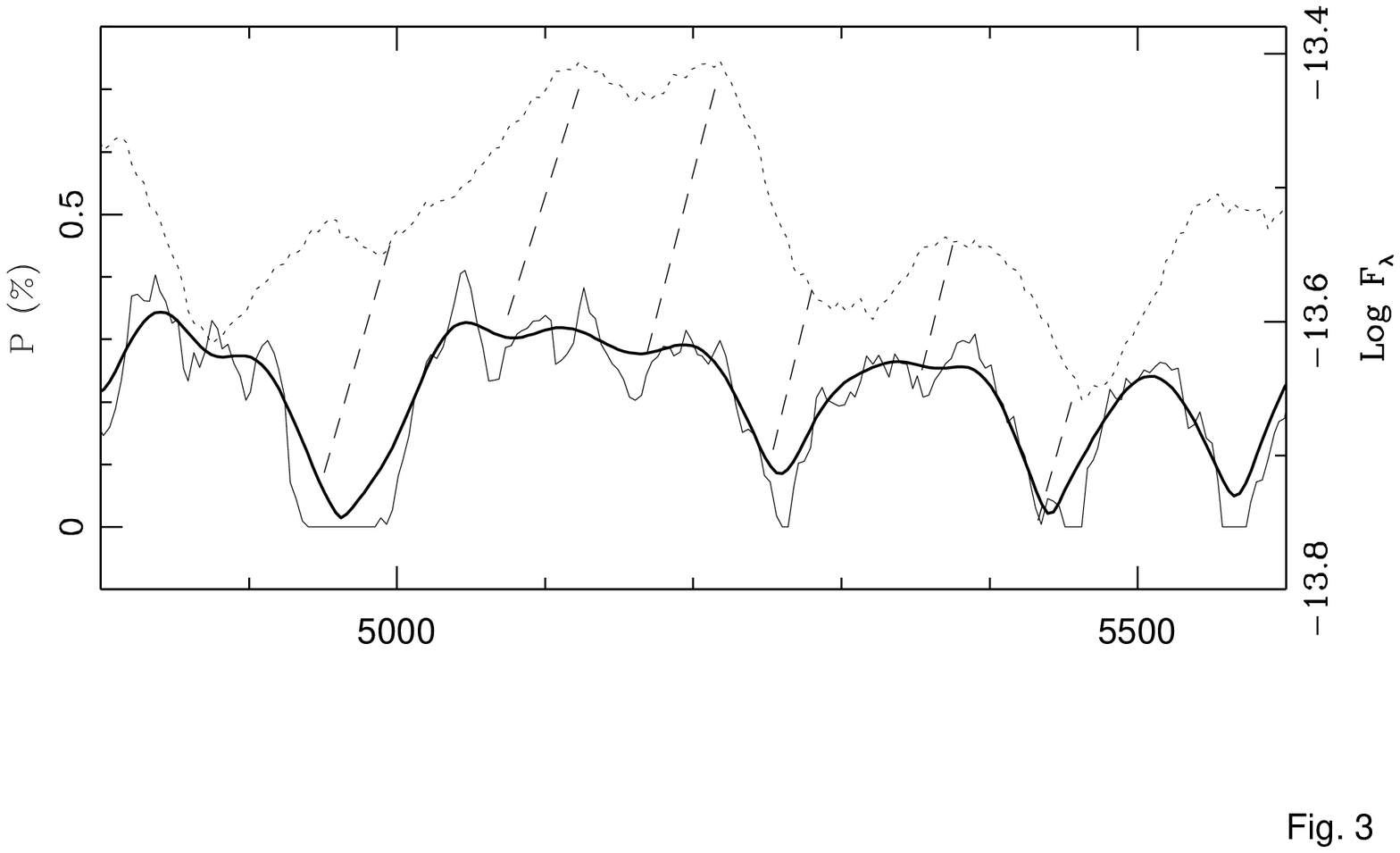,width=7.8cm,clip=,angle=0}
 \caption{
 The wavelength range showing correlations between the flux
             spectrum (dotted line), the polarization spectrum from the 
             resampled
             data (thin solid line), and the polarization of the data smoothed 
             by convolving with a Gaussian of 
             FWHM of 64.8\AA (thick solid line).   
}
 \end{figure}

The data were taken with clear and dark sky. Several unpolarized and polarized 
standard stars were observed each night. The data reduction process is the 
same as outlined in Miller, Robinson, \& Goodrich (1988). The 
instrumental polarization obtained by observing several unpolarized standards
is shown in Figure 1 (a) and 1 (b) and is found to be less than 0.1\% in the 
entire wavelength range. The polarized standards were used to calibrate the 
polarization angles. In addition, observations were also obtained by inserting 
a polarizer at the top of the waveplate to characterize the efficiency of the 
waveplate and the wavelength dependence of the fast axis of the waveplate. The
polarization efficiency is typically 95\% at around 6500 \AA\ and decreases 
to about 90\% at 4500 \AA. The variation of the fast axis with wavelength 
is less than 2$^\circ$. These effects and also the instrumental polarization 
were taken into account and corrected for in the data that will be presented 
below. 

\section{Data and Results}

Although the exact phase of the supernova awaits the publication of an 
accurate light curve, the supernova was at pre-maximum phase on April 14,
as indicated by the flux spectrum shown in Figure 2 (a). The continuum was 
very blue and the Fe II lines were still not fully developed. The Stokes 
parameters $Q$ and $U$ derived from the spectropolarimetry are shown in
Figures 2 (b) and 2 (c), and their errors in Figure 2 (d). The dominant 
error in $Q$ and $U$ was from photon statistics, and was 
the largest in the red portion of the spectra where it reached  
$1\%$; typical errors were less than 0.6\% in regions around 5500 \AA. 

Polarizations higher than 1\% can be reliably excluded from the $Q$ and $U$ 
spectra shown in Figure 2. However, at a lower level of about 0.3\%, 
some broad features are present. For a simple test, the $Q$ and $U$ 
spectra were convolved with a Gaussian of FWHM of 65 \AA. 
As shown in Figure 3, in the wavelength range from 4900 to 5500 \AA, 
there exists a correlation between the wavelengths 
of these broad features and the spectral features in the flux spectrum.

To investigate the authenticity of this polarization variation with 
wavelength, the data were resampled to increase the signal to noise ratio.  
This was done by deriving the Stokes parameters within a certain wavelength 
window $\Delta\lambda$ centered on a particular wavelength, $\lambda$. This is 
equivalent to calculating the following weighted mean of the original 
Stokes parameters: 
{\small
$$ Q(\lambda|\Delta\lambda)\ = \ 
\int_{\lambda-\Delta\lambda/2}^{\lambda+\Delta\lambda/2}\,N(\lambda^\prime) 
Q(\lambda-\lambda^\prime) d\lambda^\prime/
\int_{\lambda-\Delta\lambda/2}^{\lambda+\Delta\lambda/2}\,N(\lambda^\prime)
d\lambda^\prime,
\eqno(1)
$$}
where $N$ is the number of counts per wavelength interval.
The corresponding equation for $U$ can be similarly derived. Such an 
operation increases the signal to noise ratio of the data, but at the
expenses of degrading the spectral resolution. Fortunately, the spectral
features in a supernova spectrum are in general as broad as 150 \AA, and 
useful information can still be obtained even if the resolution is as low as 
100 \AA. The Stokes parameters with an integration window of width 
$2\times\Delta\lambda$ = 64.8 \AA\ are shown in Figure 2 (b) and (c). The 
noise 
level ($\sigma$) which is also shown in Figure 2 (d), is now typically 
around 0.1\%. 

The degree of polarization $P$ can be constructed from the Stokes parameters 
$Q$ and $U$. However, it is well known that for relatively high 
noise level, $P\ = \ \sqrt{Q^2\ +\ U^2}$ is biased toward values larger 
than the true degree of polarization \Po; $P$ is usually not the best estimate 
of \Po. 
The distribution function for $(P, \theta)$ takes the form (Simmons \& Stewart 1985)
{\small
$$
f(P, \theta)\ =\ {P\over2\pi\sigma} \exp\{-{1\over2}[(
{P\over\sigma})^2+({P_{\rm o}\over\sigma})^2-{P\over\sigma} 
{P_{\rm o}\over\sigma}
\cos(\theta-\theta_{\rm o})]\}, \eqno(2)$$}
where $(P_{\rm o}, \theta_{\rm o})$ stands for the true degree of polarization 
and polarization position angle. 
Simmons \& Stewart (1985) constructed 
optimal estimates of degree and position angle of polarization 
$(P_{\rm o}$ and  $\theta_{\rm o})$ from the marginal 
distribution of polarization obtained by integrating over $\theta$ in 
equation (2). This method, however, does not usually simultaneously optimize
$P_{\rm o}$ and $\theta_{\rm o}$. A simpler method can be constructed by 
requiring the best estimate of $(P_{\rm o}, \theta_{\rm o})$ to take the 
values for which $(P,\theta)$ gives the maximum of the distribution 
$f(P,\theta)$. It is easy to derive from equation (2) that this results in 
the following simple formula for 
\Po and $\theta_o$
$$
\begin{array}{lll}
P_{\rm o}\ &=\ P-\sigma^2/P & P\ \ge\ \sigma\nl
P_{\rm o}\ &=\ 0     & P\ <\ \sigma \nl
\theta_{\rm o} &=\  \theta & \nl
\end{array}
\eqno(3)$$
Both the polarization \Po derived from equation (3) and $P$ are shown in 
Figure 4(a). 
They differ mainly in wavelength regions with low signal to noise ratios. 
The polarization position angles could not be reliably determined for 
data with large errors, therefore only position angles calculated from the 
resampled $Q$ and $U$ are shown (Figure 4(b)).

An enlarged version of the part of the spectrum with strong correlations
between the spectral features in the polarimetry and flux spectrum is 
shown in Figure 3, where the features are marked by dashed lines for 
illustrative purposes. The corresponding spectral features in the polarization 
spectrum which are correlated with the spectral lines in the flux spectrum are 
blue shifted by about 2200 \kms. No strong correlation between the flux and 
polarization spectra could be established at wavelengths longer than 
6000 \AA. This may be due to some intrinsic characteristics of the 
polarization of Type Ia SN (see \S 4 for a more detailed discussion). 
The position angles change across spectral features. This should not be a 
surprise when compared with the best observed supernova, SN 1987A, where the 
intrinsic component shows dramatic variations of both degree and polarization 
position angle across spectral features (Cropper et al. 1988). 

\begin{figure}[h]
\psfig{figure=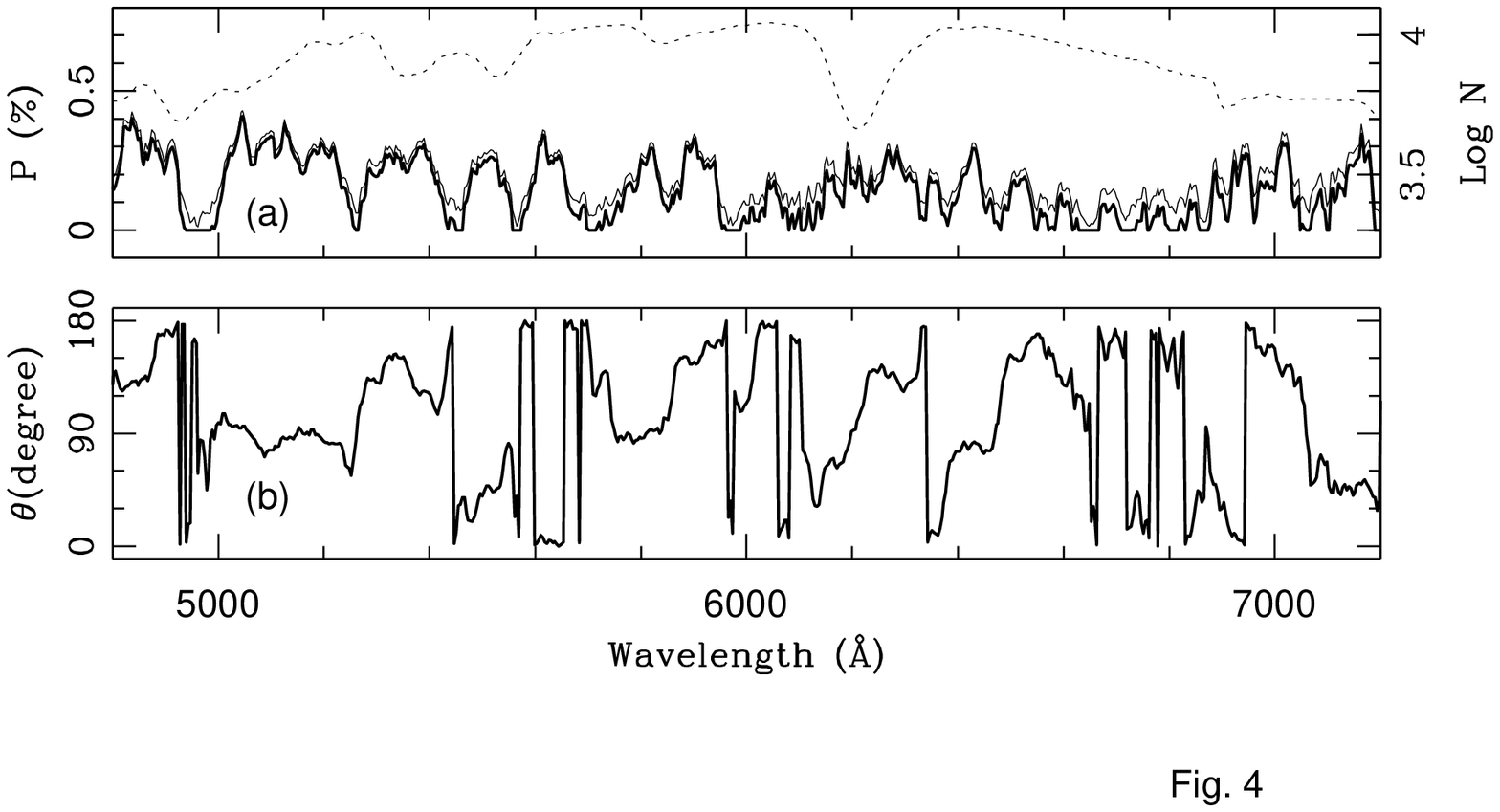,width=8.4cm,clip=,angle=0}
 \caption{
 (a) The number of counts in a single beam of a 20 minute exposure on 
             a logarithmic scale (dotted line); the degree of 
             polarization  resampled with 
             a sample window width of 64.8\AA ; the degree of polarization \Po as defined
             in equation (2) (thick solid line). (b) 
             The polarization position angle of the resampled data.  
}
 \end{figure}
 
It is impossible to accurately correct for the effect of ISP.
The Galactic extinction to the host galaxy NGC 5061 is $A_B\ = \ 0.25$, which 
alone can produce polarizations as large as 0.7\% according to the limit 
derived by Serkowski (1970). The ISP correction is even more complicated 
by the polarization due to dust in the host galaxy; however, the ISP follows 
Serkowski's law, and is a smooth function of wavelength; the uncertainties 
introduced by the ISP will not change the $Q$ and $U$ spectral structures on 
wavelength scales less than a few hundred \AA. The ISP correction can, 
however, dramatically change the appearance of the polarization spectra. The 
spectral features in Figure 4(a) may change from `emission' to `absorption' 
even if a very small amount of ISP correction $\sim 0.2\%$ is required. This 
should be borne in mind when interpretating the polarization spectra.

In addition to the spectropolarimetry discussed above, we have also obtained
broad-band polarimetry using a wide filter with pass band from 
4000 -- 7500 \AA. Three and five data sets were obtained on April 14 and May 22,
respectively. Each data set consists of four 100 second exposures which 
gave estimates of the Stokes parameters. The measured $Q$ and $U$ are 
practically zero for all data sets. The averaged 
values are $Q\ = \ 0.079\ \pm \ 0.041 \%$ and $U\ = \ 0.056\ \pm\ 0.039\%$
on April 14, and $Q\ = \ 0.090\ \pm\ 0.021\%$ and $U\ = \ 0.063\ \pm\ 0.019\%$
on May 22, where the errors are due to photon statistics. The systematic 
instrumental uncertainty is around 0.06\% 
(as indicated by repeated observations of unpolarized standards) and is 
therefore the dominant error for the broad-band data.
The broad-band data are consistent with zero polarization.
We do not confirm the short-term temporal variations of polarization
reported in early polarimetry of another Type Ia SN 1972E (Lee, Wamsteker, \& 
Wisniewski 1972, quoting observations made by Serkowski and Wamsteker).

It should be pointed out that the null detection in broad-band polarimetry 
did not conflict with the spectral features detected in the spectropolarimetry
data. As a check for consistency, we have weighted the spectropolarimetry
data with the transmission curve of the filter and the photon counts at
different wavelengths to simulate broad-band polarimetry. The resulting 
$Q$ and $U$ are in agreement with the broad-band polarimetry data.
 
\section{Models}
 
Besides the attempt to identify features of the polarization with those
of the flux spectrum, a more direct approach is a comparison with
theory which provides additional information on the spectral patterns to
be expected. A simplified approach is suitable to answer the following 
questions: what does the polarization spectrum look like for a SN~Ia,
what asphericity is needed to be consistent with the observation of SN1996X, 
and how do the predicted spectra compare with the observations ? Delayed 
detonation models (Khokhlov 1991) have been found to provide a good 
representation of the observed light curves and spectra of normal bright 
SN~Ia (H\"oflich 1995b, H\"oflich \& Khokhlov 1996 and references therein) 
and we use such a model to produce a representative theoretical SN~Ia 
polarization. 
The polarization spectrum is given by Monte Carlo calculations 
based on the following assumptions: a) homologously expanding, ellipsoidal 
envelopes with density and chemical profiles given by a hydrodynamical model,
b) occupation numbers given by local thermodynamical equilibrium (LTE); c) 
electron scattering, bound-free and free-free for continuum opacities; d) 
lines treated in a Sobolev approximation with an assumed constant 
thermalization fraction; e) line transitions result in depolarization; f) 
the temperature structure is given; g) and elliptical geometry with a axis 
ratio being independent of radius. Note that discrepancies between predicted 
and observed wavelengths of polarization features corresponding to 1000 to 2000 
\kms\ may be expected because this difference is well within the observed 
variation of expansion velocities among normal bright SNe~Ia. Lines of the same 
ion must, nevertheless, be expected to be shifted consistently.
For more details, see H\"oflich et al. (1995).
\begin{figure}[h]
\vskip -5.1cm
\vskip +5.1cm
\psfig{figure=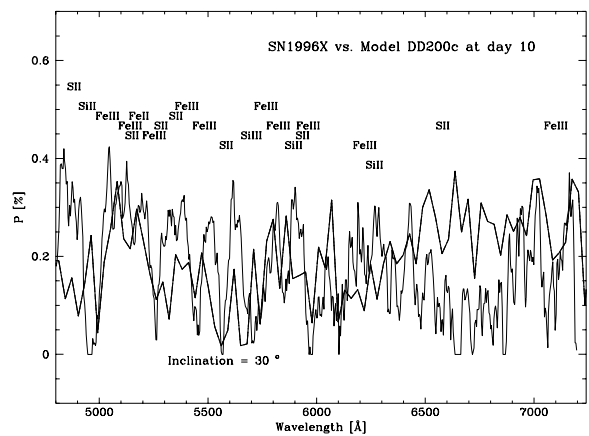,width=8.4cm,rwidth=4.4cm,angle=270}
 \caption{
 Comparison of the theoretical polarization spectrum of the 
             delayed detonatinon model DD200c five days before maximum
             light (thick line;
             H\"oflich et al. 1996; axis ratio 0.88 seen at an inclination of 
             30\degree) with the observations (thin line).
             The identification of some features are 
             given.  Most of the weaker features are due to
             iron group elements in the second and third ionization stage.
             The  S II and Si II features at about 4900, 5300, 5600 and
             5670 \AA\
             are too strong and the overall polarization above 6500 \AA\
             is too large. Possible solutions are discussed in the text.
             Note, that the observed spectrum
             is oversampled although smoothed with a triangle of 64.8 \AA
             FWHM.
}
 \end{figure}

For the  density and chemical structure we used model DD200c with T(r) based 
on light curve calculations corresponding to about 5 days before maximum 
(H\"oflich et al. 1996). In Figure 5, the  polarization spectrum for an oblate 
ellipsoid with an axis ratio of 0.89 is given which is seen at an inclination 
of 30\degree\ which fits best. Note that, quantitatively, the spectrum depends 
critically on the inclination. The depolarization in the ubiquitous metal
lines has two main effects. Firstly, the mean polarization is about a 
factor of 2 to 3 smaller compared to that of a Thomson scattering
atmosphere which provides a good approximation to Type II supernovae.
Consequently, the same asphericity will produce much less polarization for 
an SN Ia compared to an SN II. Secondly, $P$ varies strongly with wavelength 
on a typical scale of about the Doppler width. The variations with wavelength 
are significantly stronger than those in the flux spectra because it takes 
only one interaction to depolarize but many to thermalize a photon.
  
 A comparison with the observations shows some qualitative similarity  
of the wavelength distribution of the features. The pattern in the models 
is mainly influenced by the atomic physics rather than the detailed 
explosion model. Therefore, the agreement or disagreement may be used to ask 
if the measurement can be regarded as a real detection. Most of the observed 
features have their equivalents in the theoretical model, although the 
predicted polarization above 6500 \AA\ is much higher than observed (see below).
The mean size of $P$ implies an asphericity of $\approx $ 11 \% 
at an inclination angle of 30\degree for the symmetry axis of the ejecta 
(Figure 5).
 
 There are also some distinct shortcomings in the model polarization  
spectrum. The features corresponding to the products of partial 
burning (S II and Si II) are somewhat too strong compared to the
observations.  In principle, this may be compensated by slightly higher 
temperatures, small changes in the explosion model (e.g. a slightly larger
Ni production) or, as test calculations have shown, by making the
outer Si-rich layers almost spherical. A more severe problem is the spectrum 
above 6500 \AA. Although the spectral features resemble the models, the absolute
polarization is much lower. A component due to interstellar polarizaton
(U=-0.1 \%, Q = 0.25 \%) helps to bring up $P$ above 6500~\% but completely
destroys any reasonable fit at other wavelengths and, thus, does not
solve the problem. This ISP is also inconsistent with broad-band data (see
above). The increasing predicted polarization is insensitive to the details of
the explosion models. It is caused primarily by the decreasing line blending 
towards longer wavelengths. Lower line blending also implies that the 
polarization in the red is produced at deeper layers (H\"oflich 1995b). 
Consequently, a possible solution to the problem is a nearly spherical, 
inner Ni region corresponding to expansion velocities $\approx 8000$\kms. 
This, together with the depolarization in Si and S, imply that asphericities 
must be mainly attributed to the distribution of chemical elements in the
transition region between the Ni and Si layers (H\"oflich et al. 1996).

 The correlation between flux and polarization spectra depends 
sensitively on the size of the line blanketing, thermalization 
parameter of lines and the actual wavelength.  For a more detailed 
discussion see H\"oflich, Wheeler, \& Wang (1997).
 
\section{Discussions and Conclusions}

SN 1996X is the first Type Ia supernova for which 
spectropolarimetry has suggested  
a polarized component intrinsic to the supernova. 
The peak to valley spectral variation in the $Q$ and $U$
spectra (Figure 4) is as high as 0.6\% in many cases which, at an error level
of 0.15\%, gives a 4$\sigma$ detection of the spectral features. The degree 
of the intrinsic polarization can be measured in terms of the deviation of 
the spectral features from the wavelength averaged degree of polarization. 
For SN 1996X, this is about 0.3\%, which is 
small compared with the typical degree of polarization in SN 1987A and other 
Type II SNe (Wang et al. 1996). Previous observations
of SN 1983N and SN 1992A have given only upper limits on the degree of 
polarization of Type Ia supernova (Spyromilio \& Bailey, 1992; 
McCall et al. 1984). The SN 1996X observations were obtained prior to the 
optical maximum, and the detected polarization is most likely to be due to a 
distorted photosphere or element distribution.
The correlation of features  between the polarization spectra and flux 
spectra is not perfect in the entire observed wavelength range. This is
expected because such a correlation depends sensitively on multiple 
scattering effects, the ratio between continuum and line optical depths, 
and thermalization in lines. The 2200 \kms\ blue shift of the features
in the polarization spectrum relative to the flux spectrum 
(cf. Figure 3) can also be understood as due to multiple scattering.
The model calculation shows that a strong feature in the polarization spectra 
does not require the presence of a strong feature in the flux spectra. Lines 
which hardly change the flux may produce significant depolarizations.
The comparison with theoretical polarization spectra suggests an
asphericity of about 11 \% in the distribution of chemical elements.
 There are evidences that, the strong asphericity is limited to
the transition region between nuclear statistical equillibrium (NSE)
and partial burning. The distorted nature of the supernova envelope might 
arise from inhomogenous burning during the deflagration phase of the burning 
front (Khokhlov 1995).

 Such a low degree of polarization probably does not seriously endanger the 
use of Type Ia as calibrated candles for distance indicators, but nonetheless 
poses a source of uncertainty. We will discuss these effects including the 
effects of inclination and the relation between flux and polarization spectra 
quantitatively in a separate study (H\"oflich, Wheeler, \& Wang 1997).

We thank Gary Hill, and Paul Shapiro for many discussions and help. 
We are grateful to the McDonald staff, especially to David Doss and Jerry 
Martin for their excellent support. This research is supported in part by NSF 
grant AST 9528110, and NASA grant GO-2563 through the Space Telescope Science 
Institute and by  grand Ho 1177/2-1 of the Deutsche Forschungsgemeinschaft.

\end{document}